\documentclass[%
 reprint,
 amsmath,amssymb,
 aps,
]{revtex4-2}

\usepackage{graphicx}
\usepackage{dcolumn}
\usepackage{bm}
\usepackage{graphics}
\usepackage{float}

\newtheorem{theorem}{Theorem}[section]

\newenvironment{Even Case}[1][Even Case]{\begin{trivlist}
		\item[\hskip \labelsep {\bfseries #1}]}{\end{trivlist}}
\newenvironment{Odd Case}[1][Odd Case]{\begin{trivlist}
		\item[\hskip \labelsep {\bfseries #1}]}{\end{trivlist}}

\newcommand{\qed}{\nobreak \ifvmode \relax \else
	\ifdim\lastskip<1.5em \hskip-\lastskip
	\hskip1.5em plus0em minus0.5em \fi \nobreak
	\vrule height0.75em width0.5em depth0.25em\fi}
	
\begin{document}

\preprint{APS/123-QED}

\title{Verifying the surjective relation between symmetric potential function and its Scattering Matrix in 1D
}
\author{Youngik Lee}
\affiliation{Brown University, Department of Physics, Box 1843, 182 Hope Street, Barus \& Holley 545, Providence, RI 02912, USA}
\begin{abstract}
We focus on the possibility of the surjective relation between symmetric potential function and its scattering matrix in 1-dimension. The theory bases on the property of wave function symmetry and boundary conditions. This research shows the surjective relation in some particular cases, delta function potential, and finite square wall potential, and disproves the injective relation of the arbitrary potential function and its S-matrix.


\end{abstract}

\maketitle
\tableofcontents
\section{Introduction}

The scattering matrix is the matrix that relates the initial state and the final state of a physical system undergoing a scattering process \cite{1}.
Particularly this research deals with 1D symmetric potential functions and proves the surjective relation between symmetric potential function and its scattering matrix in 1-dimension and disproves the injective relation between them.
Therefore we showed that the changes of the propagated wave amplitude ratio indicate the potential changes.

First, we prove the one-to-one relation between potential and wave function, without considering phase and with considering phase.
Second, we prove the surjective relation between wave function and the scattering matrix. And disprove injective relation between scattering matrix and wave function.

\hfill

\begin{figure}[H]
	\centering
	\includegraphics[angle=0, width=8.3cm, height=2.3cm]{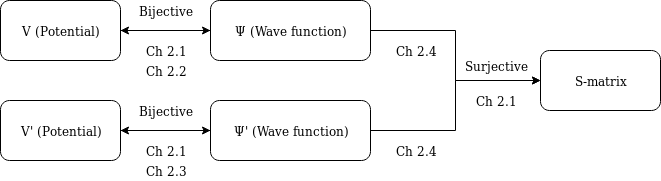}
	\caption{\label{fg1} The whole picture for the prove.}
\end{figure}

\section{Theory}

When a wave propagates through arbitrary potential, its amplitudes change. And Scattering matrix can interpret these changes by 2$\times$2 matrix in 1D.

\begin{figure}[H]
\centering
\includegraphics[angle=0, width=8.5cm, height=4.6cm]{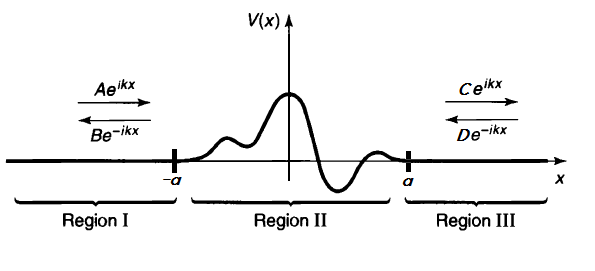}
\caption{\label{fg2} The arbitrary potential located at $-a<x<a$, and available wave functions by region. \cite{1}}
\end{figure}

Figure. \ref{fg2} shows that the arbitrary potential at $-a<x<a$ in 1D. If we can divide this situation into three regions like Fig. \ref{fg2} and $A, B, C, D$ as the wave amplitude, we can construct the scattering matrix(S-matrix) like Eq. (\ref{eq1}).

\begin{equation}\label{eq1}
\binom{B}{C}=\begin{pmatrix}
S_{11} & S_{12}\\ 
S_{21} & S_{22}
\end{pmatrix}\binom{A}{D}
\end{equation}

According to the time-independent Schrodinger equation, Eq. (\ref{eq2}), we can construct a wave equation like Eq. (\ref{eq3})$\sim$Eq. (\ref{eq5}).

\begin{equation}\label{eq2}
\left[-{\frac{\hbar ^{2}}{2m}}\frac{\partial^2}{\partial x^2}+V(\mathbf{x})\right]\Psi (\mathbf{x})=E\Psi (\mathbf{x})
\end{equation}

For the Region I, $V(x)=0$
\begin{equation}\label{eq3}
\Psi_1(x)=A e^{ikx} + B e^{-ikx}
\end{equation}

For the Region III, $V(x)=0$
\begin{equation}\label{eq4}
\Psi_3(x)=C e^{ikx} + D e^{-ikx}
\end{equation}

For the Region II,
\begin{equation}\label{eq5}
\left[-{\frac{\hbar ^{2}}{2m}}\frac{\partial^2}{\partial x^2}+V(\mathbf{x})\right]\Psi_2(\mathbf{x})=E\Psi_2(\mathbf{x})
\end{equation}

Here, $k = \sqrt{2mE}/\hbar$, and $E>0$ (Scattering condition)

\begin{theorem}\label{th1}
	\emph{Existence and uniqueness of second order differential equation}

	If the functions p(x), q(x), and g(x) are continuous on the interval $\mathbf{I}$: $\alpha<x<\beta$ and t, $t_0$ $\in$ $\mathbf{I}$, then there exists a unique solution y = $\phi(t)$, throughout the interval $\mathbf{I}$, which satisfies Eq. (\ref{eq6}).
	\begin{equation}\label{eq6}
	y'' + p(t)y' + q(t)y  =  g(t), \quad  y(t_0)  =  y_0, \quad  y'(t_0)  =  y_0'
	\end{equation}
\end{theorem}

By the Theorem. \ref{th1}, if $V(x)$ is continuous on $(-a,a)$ and for some $x_0\in (-a,a)$, $\Psi_2(x_0)$ and $\Psi_2'(x_0)$ are exist, then $\Psi_2(x)$ is also uniquely exist.

When we use the continuity condition of $\Psi(x)$ and $d\Psi(x)/dx$ (except infinite potential points) at $x=-a$ and $x=a$, we can get Eq. (\ref{eq7}) and Eq. (\ref{eq8}).

\begin{equation}\label{eq7}
\begin{cases}
A e^{ikx} + B e^{-ikx}= \Psi_2(-a)& \text{ }\\ 
& \\
A e^{ikx} - B e^{-ikx}= \frac{1}{ik}\Psi'_2(-a)& \text{ } 
\end{cases}
\end{equation}

\begin{equation}\label{eq8}
\begin{cases}
C e^{ikx} + D e^{-ikx}= \Psi_2(a)& \text{ }\\ 
& \\
C e^{ikx} - D e^{-ikx}= \frac{1}{ik}\Psi'_2(a)& \text{ } 
\end{cases}
\end{equation}

Consequently, we can compute the amplitude of the wave function as Eq. (\ref{eq9}).

\begin{equation}\label{eq9}
\begin{cases}
A=\frac{e^{ika}}{2}\left ( \Psi_2(-a)+\frac{1}{ik}\Psi_2'(-a) \right ) & \text{ } \\ 
B=\frac{1}{2e^{ika}}\left ( \Psi_2(-a)-\frac{1}{ik}\Psi_2'(-a) \right ) & \text{ } \\ 
C=\frac{1}{2e^{ika}}\left ( \Psi_2(a)+\frac{1}{ik}\Psi_2'(a) \right ) & \text{ } \\ 
D=\frac{e^{ika}}{2}\left ( \Psi_2(a)-\frac{1}{ik}\Psi_2'(a) \right )& \text{ } 
\end{cases}
\end{equation}

According to definition of S-matrix, Eq. (\ref{eq1}), we can get Eq. (\ref{eq10}).

\begin{equation}\label{eq10}
\begin{cases}
B=S_{11}A+S_{12}D& \text{ }\\ 
C=S_{21}A+S_{22}D& \text{ } 
\end{cases}
\end{equation}

\subsection{Prove Surjectivity}

And because of the $V(x)$ symmetric property, under the switching $x\leftrightarrow -x$, $A\leftrightarrow D$, $B\leftrightarrow C$, the S-matrix has to be conserved.
Therefore $S_{11}=S_{22}$, $S_{21}=S_{12}$.
Using these conditions and reorder the Eq. (\ref{eq10}), we can get Eq. (\ref{eq11}) and Eq. (\ref{eq12}).

\begin{equation}\label{eq11}
S_{11}=\frac{AB-CD}{A^2-D^2}\;, \quad S_{12}=\frac{AC-BD}{A^2-D^2}
\end{equation}

\begin{equation}\label{eq12}
S=\frac{1}{A^2-D^2}\begin{pmatrix}
AB-CD & AC-BD\\ 
AC-BD & AB-CD
\end{pmatrix}
\end{equation}

And each A, B, C, D can be express as wavefunction $\Phi$ like Eq. (\ref{eq9}).
Therefore for some symmetric continuous potential $V(x)$, if $V(x)$ has no infinite value on $[-a, a]$, then $V(x)$ and its S-matrix have the surjective relation. \qed

\subsection{one-to-one relation between $V(x)$ and $\Psi (\mathbf{x})$, without phase-shift}

According to the theorem of the uniqueness of the second-order differential equation, we know there is $\Psi (\mathbf{x})$ which satisfies Eq. (\ref{eq2}) with $V(x)$. So we can say $V(x)$ is surjective on $\Psi (\mathbf{x})$.
And to show one-to-one relation, now we need to show an injective relation between $V(x)$ and $\Psi (\mathbf{x})$.

According to Eq. (\ref{eq2}), if arbitrary symmetric potential $V(x)$ is given, then we can construct $V(x)$ as Eq. (\ref{eq13}).

\begin{equation}\label{eq13}
V(x)=E+\frac{1}{\Psi (\mathbf{x})}\left[{\frac{\hbar ^{2}}{2m}}\frac{\partial^2\Psi (\mathbf{x})}{\partial x^2}\right]
\end{equation}

But if $\Psi (\mathbf{x})=0$, Eq. (\ref{eq13}) has singularity.

So for these cases, we use Inverse Scattering Theory which can reconstruct the potential from the wave functions, and also for the function which has phase-shifts.

\subsection{one-to-one relation between $V(x)$ and $\Psi (\mathbf{x})$, with phase-shift}

The second case we need to consider is the wave function which has a phase-shift. And in this section during developing inverse scattering theory, one need the prerequisite concept of the integral equation and spectral kernels. \cite{2}

\subsubsection{Inverse Scattering Theory}

In the quantum theory of scattering normally one calculates phase shift form the potential. But by using the inverse scattering method one can reconstructing the potential from the given phase shifts. And the value called, Fredholm Determinants do an important role in the Inverse Scattering Theory.
Fredholm determinants occur also in the theory of inverse scattering. 

We can write Fredholm determinant like Eq. (\ref{eq14}) \cite{2}

\begin{equation}\label{eq14}
F_+(t)=\prod_{n=0}^{\infty}(1-\lambda_{2n}(t)), \quad F_-(t)=\prod_{n=0}^{\infty}(1-\lambda_{2n+1}(t))
\end{equation}

where $\lambda_n$ are the eigenvalues of the integral equation

\begin{equation}\label{eq15}
\lambda f(x)=\int^{1}_{-1} \frac{\sin(x-y)}{(x-y)\pi} f(y) dy
\end{equation}

Here $\lambda$ arranged in descending order

\begin{equation}\label{eq16}
1\geq\lambda_0\geq\lambda_1\geq\cdots\geq0
\end{equation}

And because of the periodicity of trigonometric functions, it is convenient to parametrize x with $\pi$t. So when we choose new variables $\xi = x\pi t$ , $\eta = y\pi t$, we can rewrite Eq. (\ref{eq15}) as Eq. (\ref{eq17})

\begin{equation}\label{eq17}
\lambda f\left(\frac{\xi}{\pi t}\right)=\frac{1}{\pi}\int^{\pi t}_{-\pi t} \frac{\sin(\xi-\eta)}{\xi-\eta} f\left(\frac{\eta}{\pi t} \right)d\eta 
\end{equation}

And as $f(-x)=\pm f(x)$, the $\lambda_j$ are the eigenvalues of the integral equations Eq. (\ref{eq19}), Eq. (\ref{eq20}) with kernels, Eq. (\ref{eq18})

\begin{equation}\label{eq18}
K_{\pm}(x,y)=\frac{1}{\pi}\left(\frac{\sin(x-y)}{x-y}\pm \frac{\sin(x+y)}{x+y} \right)
\end{equation}

\begin{equation}\label{eq19}
\lambda_{2j}g(x)=\int^{\pi t}_{0} K_+(x,y)g(y)dy
\end{equation}

\begin{equation}\label{eq20}
\lambda_{2j+1}g(x)=\int^{\pi t}_{0} K_-(x,y)g(y)dy
\end{equation}

Therefore $F_{\pm}(t)$ are the Fredholm determinants of the kernels Eq. (\ref{eq18})  ( or Eq. (\ref{eq22}), Eq. (\ref{eq23})) over the interval (0, $\tau$). So we can rewrite Fredholm determinant as Eq. (\ref{eq21})

\begin{equation}\label{eq21}
F_j(\tau)=det[1-K_j]^{\tau}_{0}, \quad \tau=\pi t
\end{equation}

\begin{equation}\label{eq22}
K_{+}(x,y)=\frac{2}{\pi}\int^{1}_{0}\cos(kx)\cos(ky)dk
\end{equation}

\begin{equation}\label{eq23}
K_{-}(x,y)=\frac{2}{\pi}\int^{1}_{0}\sin(kx)\sin(ky)dk
\end{equation}

In the inverse scattering theory we are dealing with two potential $V_1(x)$, $V_2(x)$ on the half-line $0<x<\infty$, and $V_1$ being supposed unknown and $V_2$ known. 
We suppose $V(x)$ is symmetric, so the potential range $0<x<\infty$ is valid.

We have two corresponding families of wave-functions $\Psi_1(\mathbf{k,x})$, $\Psi_2(\mathbf{k,x})$ satisfying the wave equations Eq. (\ref{eq24})

\begin{equation}\label{eq24}
\left[\frac{\partial^2}{\partial x^2}+k^2-W_j(x)\right]\Psi_j(\mathbf{k,x})=0, \quad j=1,2
\end{equation}

Here $W_j=\frac{2mV_j}{\hbar ^{2}}, k=\frac{2mE}{\hbar ^{2}}$.

And the generating spectral kernels $K_1$, $K_2$ defined by Eq. (\ref{eq25})

\begin{equation}\label{eq25}
\delta(x-y)-K_j (x, y) =\int \Psi_j(\mathbf{k,x}) \Psi_j(\mathbf{k,y}) dk
\end{equation}

There are two methods are available for reconstructing potentials.
First is the Gel’fand–Levitan method(GL method) and second is the Marchenko method(MK method). GL method constructs the potential by working from $x=0$ upwards. And MK method works downwards from $x=\infty$. So the first method is related to a Fredholm determinant over the interval (0, $\tau$), while the second involves that over the interval ($\tau$, $\infty$). 

So by using Gel'fan-Levitan method, one needs to show that there is the potential that can be uniquely exist reconstructed by the wave function. Furthermore, by using the property that Gel'fan-Levitan method and Marchenko method are equivalent, we can calculate the form of the original potential analytically.

$\\$
\paragraph{Gel’fand–Levitan Method}
$\\$
$\\$
In the Gel’fand–Levitan method \cite{3}, the wave-functions $\Psi_u(\mathbf{k,x})$ are subject to boundary conditions at $x=0$, namely

\begin{equation}\label{eq26}
\Psi_1(\mathbf{k,0})= \Psi_2(\mathbf{k,0})= g q(k)
\end{equation}

\begin{equation}\label{eq27}
\Psi'_u(\mathbf{k,0})= h_jq(k)
\end{equation}

Here q(k) is a given function and g, $h_1$, $h_2$ are given coefficients.

The Fredholm determinants are defined on the finite interval [0, $\tau$], and satisfy the conditions Eq. (\ref{eq28}), Eq. (\ref{eq29}) \cite{4}

\begin{equation}\label{eq28}
W_1(\tau)-W_2(\tau)=2\frac{d^2}{d\tau^2} \log  \left( \frac{F_1(\tau)}{F_2(\tau)} \right)
\end{equation}

\begin{equation}\label{eq29}
h_1-h_2=g\frac{d}{d\tau} \log  \left( \frac{F_1(\tau)}{F_2(\tau)} \right)\Big|_{\tau=0}
\end{equation}

$\\$
\paragraph{Marchenko Method}
$\\$
$\\$
In the Marchenko version \cite{5} of the theory, the wave-functions $\Psi_u(\mathbf{k,x})$ are required to become asymptotically equal at infinity, thus Eq. (\ref{eq30}) is valid under uniformly in k.

\begin{equation}\label{eq30}
\Psi_1(\mathbf{k,x})-\Psi_2(\mathbf{k,x})\rightarrow 0\quad as\quad x\rightarrow\infty
\end{equation}

The potentials $W_1$, $W_2$  must approach each other closely enough so that the integral of Eq. (\ref{eq31}) converges at infinity. 

\begin{equation}\label{eq31}
\int \left |W_1(\tau)-W_2(\tau) \right | \tau d\tau
\end{equation}

Also, in this case, the Fredholm determinants Eq. (\ref{eq32}) are defined on the infinite interval [$\tau$, $\infty$], with the kernel $G_j$ and satisfy the conditions Eq. (\ref{eq33})

\begin{equation}\label{eq32}
\Delta_j(\tau)=det[1-G_j]^{\tau}_{0}, \quad \tau=\pi t
\end{equation}

\begin{equation}\label{eq33}
W_1(\tau)-W_2(\tau)=2\frac{d^2}{d\tau^2} \log  \left( \frac{\Delta_1(\tau)}{\Delta_2(\tau)} \right)
\end{equation}

Here, Eq. (\ref{eq28}) and Eq. (\ref{eq33}) relate two sets of Fredholm determinants one defined on the finite interval (0, $\tau$), and the other defined over the infinite interval ($\tau$, $\infty$).

When the Gel’fand–Levitan and Marchenko methods are applied to the inverse scattering problem, it is normal to assume that the unknown wavefunctions $\Psi_1(\mathbf{k,x})$ form a complete orthonormal set.
Then Eq. (\ref{eq24}) implies

\begin{equation}\label{eq34}
K_1=0, \quad F_1=\Delta_1=1
\end{equation}

And during the using both methods, one of the potentials is laid by the Fredholm determinant we want to study. And the second comparison potential has to satisfy the boundary conditions Eq. (\ref{eq26}) and Eq. (\ref{eq27}) near the origin, and the convergence condition Eq. (\ref{eq31}) at large distances. And also need not be the same near the origin and at large distances, and the choice might be good to be simple enough to allow computation of the corresponding Fredholm determinants.

\subsubsection{Application of Gel'fand-Levitan Method}

\paragraph{Finding $\Psi_1(\mathbf{k,x})$}
$\\$
$\\$
So now let's apply the Gel’fand–Levitan formalism to the potentials and prove we can uniquely determine $W_1$

\begin{equation}\label{eq35}
W_1(\tau)=W_{\pm}(\tau)=-2\frac{d^2}{d\tau^2} \log  F_{\pm}(t)-1, \quad \tau=\pi t
\end{equation}

Here the comparison potential is,

\begin{equation}\label{eq36}
W_{2}(\tau)=-1
\end{equation}

So to calculate the $W_1$, we should find the value of $F_{\pm}(t)$.
According to Appendix A, Eq. (\ref{eq88}), $W_{\pm}$ has asymptotic behavior of Eq. (\ref{eq37})

\begin{equation}\label{eq37}
W_{\pm}(\tau)\sim-\frac{1}{4}\tau^{-2} \quad as \quad \tau\rightarrow\infty
\end{equation}

We take $\Psi_1(\mathbf{k,x})$ to be a complete orthonormal set of solutions of Eq. (\ref{eq24}), so that Eq. (\ref{eq28}) holds with

\begin{equation}\label{eq38}
F_1(\tau)=1,\quad F_2(\tau)=F_{\pm}(\tau)
\end{equation}

To satisfy Eq. (\ref{eq25}), the wave-functions $\Psi_2(\mathbf{k,x})$ must be cosines and sines of $(k^2+1)^{1/2}x$. We have to normalize these wave-functions so that

\begin{equation}\label{eq39}
K_2(x,y)=K_{\pm}(x,y)
\end{equation}

with $K_2$ given by Eq. (\ref{eq25}). 

And $K_\pm$ by Eq. (\ref{eq22}), Eq. (\ref{eq23}), we get Eq. (\ref{eq40}) which is satisfied by choosing Eq. (\ref{eq41}), Eq. (\ref{eq42})

\begin{equation}\label{eq40}
\int^{\infty}_{0} \Psi_2(\mathbf{k,x}) \Psi_2(\mathbf{k,y}) dk = \frac{2}{\pi} \int^{\infty}_{1} {^{\cos}_{\sin}kx}{^{\cos}_{\sin}ky} dk
\end{equation}

\begin{equation}\label{eq41}
\Psi_2(\mathbf{k,x})=A(k){^{\cos}_{\sin}((k^2+1)^{1/2}x)}
\end{equation}

\begin{equation}\label{eq42}
A(k)=\left( \frac{2}{\pi} \right)^{1/2} \left[\frac{k^2}{k^2+1} \right]^{1/4}
\end{equation}

We next have to determine the boundary conditions satisfied by $\Psi_1(\mathbf{k,x})$ and $\Psi_2(\mathbf{k,x})$ at $x=0$. 

\begin{Even Case}
	\emph{when $V_1=W_+$}
	
	In the even case, when $V_1=W_+$, Eq. (\ref{eq26}) and Eq. (\ref{eq27}) hold with
	\begin{equation}\label{eq43}
	q(k)=A(k), \quad g=1, \quad h_2=0
	\end{equation}
	In this case Eq. (\ref{eq29}) gives
	\begin{equation}\label{eq44}
	h_1=-\frac{d}{d\tau}\left[\log  F_{+}(t) \right]_{\tau=0}=K_{+}(0,0)=\frac{2}{\pi}, \quad \tau=\pi t
	\end{equation}
	Therefore the boundary conditions for $\Psi_1$ are
	\begin{equation}\label{eq45}
	\Psi_1(\mathbf{k,0})=A(k),\quad \Psi'_1(\mathbf{k,0})=\frac{2}{\pi}A(k)\qed
	\end{equation}
\end{Even Case}

\begin{Odd Case}
		\emph{when $V_1=W_-$}
	
	In the odd case, when $V_1=W_-$, Eq. (\ref{eq29}) gives
	\begin{equation}\label{eq46}
	h_1=h_2
	\end{equation}

	and the boundary conditions for $\Psi_1$ are

	\begin{equation}\label{eq47}
	\Psi_1(\mathbf{k,0})=0,\quad \Psi'_1(\mathbf{k,0})=(k^2+1)^{1/2}A(k)\qed
	\end{equation}
\end{Odd Case}

Now we need to determine the behaviour of $\Psi_1(\mathbf{k,x})$ as $x\rightarrow\infty$. Because the potential $W_1$ decreases according to Eq. (\ref{eq37}) at infinity, and the $\Psi_1(\mathbf{k,x})$ are an orthonormal system, we have

\begin{equation}\label{eq48}
\Psi_1(\mathbf{k,x})\sim \sqrt{\frac{2}{\pi}}{^{\cos}_{\sin}(kx+\eta(k))}
\end{equation}

where the phase-shift $\eta(k)$ must be calculated separately for the even and odd cases.

$\\$
\paragraph{Finding phase-shift $\eta(k)$}
$\\$
$\\$
Next, let's define the value of the phase-shift $\eta(k)$, so uniquely determine the wavefunction.
Following Jost \cite{6}, we define $u(k,x)$ to be the solution of Eq. (\ref{eq24}) with potential $W_1$ and the asymptotic behavior

\begin{equation}\label{eq49}
u(k,x)\sim \exp(ikx)\quad as\quad x\rightarrow\infty
\end{equation}

The functions Eq. (\ref{eq50}) are the boundary values of a function analytic in the half-plane (${\rm Im}(k) > 0$), with the symmetry property Eq. (\ref{eq51}) and the asymptotic behavior Eq. (\ref{eq52})

\begin{equation}\label{eq50}
a(k) = u(k, 0),\quad b(k)= u'(k, 0)
\end{equation}

\begin{equation}\label{eq51}
a(-k)= a^*(k), \quad b(-k)= b^*(k),\quad k \in \mathbb{R}
\end{equation}

\begin{equation}\label{eq52}
a(k)\sim 1,\quad b(k) \sim ik,\quad k\rightarrow\infty
\end{equation}

The Wronskian Eq. (\ref{eq53}) is independent of $x$. Equating its value at $x=0$ with its value at $x=\infty$, we can get Eq. (\ref{eq54})

\begin{equation}\label{eq53}
u(-k, x)u'(k, x)-u'(-k, x)u(k, x)
\end{equation}

\begin{equation}\label{eq54}
a^*(k)b(k)-a(k)b^*(k) = 2ik
\end{equation}

And here, one can compare Eq. (\ref{eq48}) with Eq. (\ref{eq49}), one can get Eq. (\ref{eq55})

\begin{equation}\label{eq55}
\Psi_1(\mathbf{k,x}) = \sqrt{\frac{2}{\pi}}{^{Re}_{Im}\left[u(k,x)\exp(i\eta(k))\right]}
\end{equation}

\begin{Even Case}
	\emph{when $V_1=W_+$}
	
	Consider first the even case. Then the boundary conditions Eq. (\ref{eq45}) together with Eq. (\ref{eq55}) imply

\begin{equation}\label{eq56}
Re\left[\exp(i\eta(k))a(k)\right]=\left[\frac{k^2}{k^2+1}\right]^{1/4}
\end{equation}

\begin{equation}\label{eq57}
Re\left[\exp(i\eta(k))\left(b(k)-\frac{2}{\pi}a(k)\right)\right]=0
\end{equation}

And here let's define Eq. (\ref{eq58}) like below,

\begin{equation}\label{eq58}
\phi(k) = b(k)-\frac{2}{\pi}a(k)
\end{equation}

Then the Eq. (\ref{eq58}) is analytic in the upper half-plane and satisfies the same conditions Eq. (\ref{eq51}), Eq. (\ref{eq52}) as b(k). According to Eq. (\ref{eq57})

\begin{equation}\label{eq59}
\exp(i\eta(k))= i\frac{\phi^*(k)}{|\phi(k)|}
\end{equation}

When one substitute Eq. (\ref{eq59}) into (\ref{eq56}) and make use of Eq. (\ref{eq54}), we obtain

\begin{equation}\label{eq60}
|\phi(k)|=(k^2(k^2+1))^{1/4}
\end{equation}

The only analytic function satisfying Eq. (\ref{eq51}), Eq. (\ref{eq52}) and Eq. (\ref{eq60}) is

\begin{equation}\label{eq61}
\phi(k) = [-k(k + i)]^{1/2}
\end{equation}

Finally putting Eq. (\ref{eq61}) back into (\ref{eq59}), we find

\begin{equation}\label{eq62}
\exp(i\eta(k))=\left[\frac{k-i}{k+i}\right]^{1/4}
\end{equation}

and so the phase-shift is given by

\begin{equation}\label{eq63}
k \tan 2\eta(k)=-1 \qed
\end{equation}

\end{Even Case}

\begin{Odd Case}
	\emph{when $V_1=W_-$}

	In the odd case, the boundary conditions Eq. (\ref{eq47}) implies

\begin{equation}\label{eq64}
{\rm Im}[\exp(i\eta(k))a(k)]= 0
\end{equation}

\begin{equation}\label{eq65}
{\rm Im}[\exp(i\eta(k))b(k)]=[k^2(k^2 + 1)]^{1/4}
\end{equation}

And Eq. (\ref{eq64}) implies Eq. (\ref{eq66}) and Eq. (\ref{eq66}) with Eq. (\ref{eq65}) and Eq. (\ref{eq54}) gives Eq. (\ref{eq67})

\begin{equation}\label{eq66}
\exp(i\eta(k))=\frac{a^*(k)}{|a(k)|}
\end{equation}

\begin{equation}\label{eq67}
|a(k)|=\left(\frac{k^2}{k^2+1}\right)^{1/4}
\end{equation}

The analytic function $a(k)$ is then

\begin{equation}\label{eq68}
a(k)=\left(\frac{k}{k+i}\right)^{1/2}
\end{equation}

So Eq. (\ref{eq66}) gives $\exp(i\eta(k))=[(k+i)/(k-i)]^{1/4}$, or

\begin{equation}\label{eq69}
k \tan 2\eta(k)=1 \qed
\end{equation}
\end{Odd Case}

Therefore for both cases can be expresesed in the Equation

\begin{equation}\label{eq70}
\eta(k)=\mp \frac{1}{2}\arctan(k^{-1})
\end{equation}

and this shows the potentials $W_{\pm}(\tau)$ are uniquely determined \cite{7} by the property that they give scattering states Eq. (\ref{eq48}) with the phase-shifts Eq. (\ref{eq70}), and no bound states. \qed

So this shows we can uniquely construct the potential W from the wavefunction.
Therefore we proved the one-to-one relation between $V(x)$ and $\Psi(\mathbf{k,x})$.
For reconstructing the exact form of the $V(x)$ is on Appendix B

\subsection{Disprove Injectivity}

Then to disprove the injective relation, consider particular case. According to Eq. (\ref{eq9}) if values of $\Psi_2(x)$ and $\Psi_2'(x)$ at $x=a$ and $x=-a$ are equal in some different potential $V_1(x)$ and $V_2(x)$, then its S-matrixs are same by Eq. (\ref{eq9}) and Eq. (\ref{eq12}).
So first we should prove one-to-one relation between $\Psi(x)$ and $V(x)$, and find two different wave function which has same values of $\Psi_2(x)$ and $\Psi_2'(x)$ at $x=a$ and $x=-a$.

\subsubsection{Disprove the injectivity of $\Psi (\mathbf{x})$ on S-matrix}

Second, now disprove the injective relation on $\Psi (\mathbf{x})$ and S-matrix.
If we suppose the two different symmetric wave function $\Psi (\mathbf{x})$ and $\Psi_0 (\mathbf{x})$ in Region II, like Eq. (\ref{eq71}), we can get Eq. (\ref{eq73}).

\begin{equation}\label{eq71}
\Psi (\mathbf{x})=\Psi_0 (\mathbf{x})+f(x)
\end{equation}

Here, $f(x)$ is given as Eq. (\ref{eq72}) and it is also symmetric.

\begin{equation}\label{eq72}
f(x)=\begin{cases}
(x-a)^2& \text{ if } 0\leq x\leq a \\ 
(x+a)^2& \text{ if } -a\leq x\leq 0 
\end{cases}
\end{equation}

\begin{equation}\label{eq73}
\begin{cases}
\Psi'(-a)=\Psi'_0(-a) \\ 
\Psi'(a)=\Psi'_0(a) \\ 
\Psi(-a)=\Psi_0(-a) \\ 
\Psi(a)=\Psi_0(a) 
\end{cases}
\end{equation}

By Eq. (\ref{eq9}) and Eq. (\ref{eq73}), $A, B, C, D$ is same at $\Psi (\mathbf{x})$ and $\Psi_0 (\mathbf{x})$ cases. And by Eq. (\ref{eq11}), S-matrix of $\Psi (\mathbf{x})$ is same as $\Psi_0 (\mathbf{x})$. Therefore the relation between wave function to S-matrix is the surjective and not the injective. Thus, according to the one-to-one relation of $\Psi (\mathbf{x})$ and $V(x)$, the relation between $V(x)$ to S-matrix is the surjective and not the injective.

\section{Application}

Now for checking our work correctness, we are going to consider two potential cases. First is the finite square wall and the second is delta function. But delta function potential has infinite value on $x=0$, so we are going to deal with this example as an exceptional case. 

Suppose wave is moving from Region I to Region III, then D=0. Therefore $B/A=S_{11}$, $C/A=S_{21}$. 

First Finite square wall potential case, $V(x)=-V_0$ ($-a<x<a$). In this case, $\Psi(x)$ is divided into three regions like Eq. (\ref{eq74}).

\begin{equation}\label{eq74}
\begin{cases}
\Psi_1(x)=Ae^{ikx} + Be^{-ikx}& \text{for Region I}\\ 
\Psi_2(x)=F\sin(lx) + G\cos(lx)& \text{for Region II}\\ 
\Psi_3(x)=Ce^{ikx} + De^{-ikx}& \text{for Region III} 
\end{cases}
\end{equation}

Here $l$ and $k$ are in Eq. (\ref{eq75}).

\begin{equation}\label{eq75}
l=\frac{\sqrt{2m(E+V_0)}}{\hbar}, \quad k=\frac{2mE}{\hbar}
\end{equation}

When we use the Eq. (\ref{eq9}) and Eq. (\ref{eq11}) conditions with Eq. (\ref{eq74}), we can get Eq. (\ref{eq76}) and S-matrix as Eq. (\ref{eq77}).

\begin{equation}\label{eq76}
\begin{cases}
B=i\frac{\sin(2la)}{2kl}(l^2-k^2)C\\
&\\
C=\frac{e^{-2ika}}{\cos(2la)-i\frac{k^2+l^2}{2kl}\sin(2la)}A& \text{}
\end{cases}
\end{equation}

\begin{equation}\label{eq77}
S=W\begin{pmatrix}
i\frac{l^2-k^2}{2kl}\sin(2la) & 1\\ 
1 & i\frac{l^2-k^2}{2kl}\sin(2la)
\end{pmatrix}
\end{equation}

Here $W$ is given as Eq. (\ref{eq78}).

\begin{equation}\label{eq78}
W=\frac{e^{-2ika}}{\cos(2la)-i\frac{k^2+l^2}{2kl}\sin(2la)}
\end{equation}

Second delta potential case, $V(x)=-\alpha\delta(x)$ and $\alpha>0$, we can get the difference between primary differential value of $\Psi(x)$ at $x=0$ as Eq. (\ref{eq79}) \cite{1}.

\begin{equation}\label{eq79}
\Delta\left(\frac{d\Psi}{dx}\right)=-\frac{2m\alpha}{\hbar^2}\Psi(0)
\end{equation}

In delta function potential case, Region II disappears, so entire wave function $\Psi$ is constructed by two parts. $\Psi_1=Ae^{ikx}+Be^{-ikx}$ and $\Psi_2=Ce^{ikx}+De^{-ikx}$. 

Using continuous condition of $\Psi$ on $x=0$, we can show that $A+B=C+D$. Also when we consider the continuous condition of $d\Psi/dx$, we can get Eq. (\ref{eq80}). 

\begin{equation}\label{eq80}
\begin{cases}
d\Psi/dx=ik(Ce^{ikx}-De^{-ikx})& \text{for x$>$0 }\\ 
d\Psi/dx=ik(Ae^{ikx}-Be^{-ikx})& \text{for x$<$0}
\end{cases}
\end{equation}

So $d\Psi/dx\mid_{0^+}=ik(C-D)$ and $d\Psi/dx\mid_{0^-}=ik(A-B)$.

By using Eq. (\ref{eq79}), we can get Eq. (\ref{eq81}).

\begin{equation}\label{eq81}
\begin{cases}
A+B=C+D& \text{ }\\ 
ik(C-D-A+B)=-\frac{2m\alpha}{\hbar^2}(A+B)& \text{ } 
\end{cases}
\end{equation}

And like we discussed when the wave comes from Region I to III, then $D=0$. 
Therefore S-matrix of delta function potential is given by Eq. (\ref{eq82}).

\begin{equation}\label{eq82}
S=\frac{1}{1-i\beta}\begin{pmatrix}
i\beta & 1\\ 
1 & i\beta
\end{pmatrix}
\end{equation}

Here, $\beta=m\alpha/k\hbar^2$.

Therefore for some symmetric continuous potential $V(x)$, which has no infinite value on $[-a, a]$, $V(x)$ and its S-matrix have only the surjective relation. And also we showed that this statement is valid on finite square wall and delta-function potential cases. \qed

\section{Appendix}

\subsection{Asymptotics behavior of $\log F_{\pm}(t)$}

A Szegö limit theorem \cite{9} describes the asymptotic behavior of the determinants of large Toeplitz matrices.

\subsubsection{First Szegő theorem}

If $f(\theta)$ is a positive function over 0 $\leq$ $\theta$ $\leq 2\pi$, its derivative satisfies a Lipschitz condition and $F_N(f)$ is the $N\times N$ Toeplitz determinant 
	
\begin{equation}\label{eq83}
F_N(f) = det\left[\frac{1}{2\pi}\int^{2\pi}_{0}f(\theta)\exp(i(j-k)\theta)d\theta\right]_{j,k=1,...,N}
\end{equation}

Then as $N$ goes to infinity, Eq. (\ref{eq84}) is hold. \qed

\begin{equation}\label{eq84}
\log F_N(f)\approx Nf_0+\frac{1}{4}\sum^{\infty}_{k=1}kf_{k}f_{-k}+\mathcal{O}(1)
\end{equation}

Here $f_k$ are the Fourier coefficients of $\log f(\theta)$,

\begin{equation}\label{eq85}
\log f(\theta)=\sum^{\infty}_{k=-\infty}f_{k}e^{ik\theta}
\end{equation}

And Widom (1971) extended first the Szegö theorem for functions $f(\theta)$ which are positive only on an arc of the unit circle.

\subsubsection{Second Szegő theorem}

Let $f(\theta)$ = 1, if $\alpha \leq \theta \leq 2\pi-\alpha$, and $f(\theta)$ = 0 if either $\theta \leq \alpha$ or $\theta>2\pi-\alpha$. Then as $N\rightarrow\infty$, (where $\zeta'$(z) is the derivative of the Riemann zeta function.) 

\begin{align}\label{eq86}
&\log F_N(f) \approx \nonumber \\
&N^2\log\cos\frac{\alpha}{2}-\frac{1}{4}\log\left(N\sin\frac{\alpha}{2}\right)+\frac{1}{12}\log2+3\zeta'(-1)
\end{align}

Taking the limit $\alpha N = 2\pi t \gg1$, one has,

\begin{equation}\label{eq87}
\log F(t)\approx-\frac{1}{2}(\pi t)^2-\frac{1}{4}\log (\pi t)+\frac{1}{12}\log 2+3\zeta'(-1)
\end{equation}

Eq. (\ref{eq87}) with Eq. (\ref{eq89}) gives Eq. (\ref{eq88}) \qed
 
\begin{align}\label{eq88}
&\log F_{\pm}(t)\approx\nonumber \\
&-\frac{(\pi t)^2}{4}\mp\frac{\pi t}{2}-\frac{1}{8}\log(\pi t)+\left(\frac{1}{24}\pm\frac{1}{4}\right)\log 2 + \frac{3}{2}\zeta'(-1)
\end{align}

The details of this analysis of finding these terms can be found from a theorem of Widom, CloiseauxJ., Mehta, M.L.(1973) \cite{10}

Because we get a relation between the asymptotic behaviours of $F_+(t)$, $F_-(t)$ and their product 
$F(t)=F_+(t)F_-(t)$ for large t, therefore

\begin{equation}\label{eq89}
\log F_{\pm}(t) \approx \frac{1}{2}\log F(t)\mp\frac{\pi}{2}t
\end{equation}

Here when we use some analysis one can find the coefficients of $t^2$, t and $\log(t)$ in the asymptotic expansion of $\log F(t)$.

\subsection{Analytic form of the potential $V(x)$}

\subsubsection{Finding $\Psi_2(\mathbf{k,x})$ and phase-shift $\theta(k)$}

In this appendix section, one can use the Marchenko formalism \cite{2} to the potential $W_1(\tau)$ defined by Eq. (\ref{eq35}). For the first step, we should choose the comparison potential $W_2(\tau)$. And when we choose the simple potential, condition Eq. (\ref{eq90}) makes the integral Eq. (\ref{eq31}) converge. 

\begin{equation}\label{eq90}
W_{2}(\tau)\sim-\frac{1}{4}\tau^{-2}
\end{equation}

The wave-functions $\Psi_2(\mathbf{k,x})$ are now Bessel functions with quantity "$kx$", and the phase-shift produced by the potential $W_2(\tau)$ is ($\mp\pi/4$) independent of $k$. This phase-shift agrees with the Eq. (\ref{eq70}) at $k=0$, reflects that $W_1$ and $W_2$ have the same behaviour at infinity.

But one can also make a better choice for $W_2$, since The phase-shift Eq. (\ref{eq70}) behaves like Eq. (\ref{eq91}) with an error of order $k^3$ for small $k$. 

\begin{equation}\label{eq91}
\eta(k)\sim \mp \left( \frac{1}{4}\pi-\frac{1}{2}k \right)
\end{equation}

A Bessel function of the quantity $(k(x\mp1/2))$ gives the phase-shift Eq. (\ref{eq91}) for all $k$. 

\begin{equation}\label{eq92}
W_{2}(\tau)=-\frac{1}{4}\left(\tau\pm \frac{1}{2}\right)^{-2}
\end{equation}

Therefore one can choose for $W_2$ the potential Eq. (\ref{eq92}) with the expectation that this will make the difference ($W_1-W_2$) decrease much more rapidly as $\tau$$\rightarrow$$\infty$. The singularity of $W_2(\tau)$ at $\tau$ = 1/2 (in the odd case) means that the analysis is valid only for $\tau>1/2$. 

The wave-functions $\Psi_2(\mathbf{k,x})$ are solutions of the equation

\begin{equation}\label{eq93}
\left[\frac{\partial^2}{\partial x^2}+k^2+\frac{1}{4}\left(x\pm \frac{1}{2}\right)^{-2}\right]\Psi_u(\mathbf{k,x})=0
\end{equation}

with asymptotic behaviour determined by Eq. (\ref{eq30}). It is convenient to use the notations

\begin{equation}\label{eq94}
j(z)=z^{1/2}J_0(z), \quad y(z)=z^{1/2}Y_0(z)
\end{equation}

\begin{equation}\label{eq95}
h(z)=z^{1/2}H^1_0(z), \quad k(z)=z^{1/2}K_0(z)
\end{equation}

for the Bessel functions. We take for the wave-functions $\Psi_1(\mathbf{k,x})$ the same complete orthonormal set with asymptotic behavior given by Eq. (\ref{eq48}) and Eq. (\ref{eq70}). 

Then Eq. (\ref{eq30}) implies

\begin{equation}\label{eq96}
\Psi_2(\mathbf{k,x})=\alpha(k)j\left(k\left(x\pm \frac{1}{2}\right)\right) \pm \beta(k)y\left(k\left(x\pm \frac{1}{2}\right)\right)
\end{equation}

with Eq. (\ref{eq97}), Eq. (\ref{eq98})

\begin{equation}\label{eq97}
\alpha(k)=\cos\left(\frac{\theta(k)}{2}\right), \quad \beta(k)=\sin\left(\frac{\theta(k)}{2}\right)
\end{equation}

\begin{equation}\label{eq98}
\theta(k)=k-\arctan(k)
\end{equation}

\subsubsection{Analytic form of the origianl potential W}

The Marchenko Eq. (\ref{eq33}) becomes

\begin{equation}\label{eq99}
W_1(\tau)=W_{\pm}(\tau)=-2\frac{d^2}{d\tau^2} \log  \Delta_{\pm}(t)-\frac{1}{4}\left(\tau \pm \frac{1}{2}\right)^{-2}
\end{equation}

\begin{equation}\label{eq100}
\Delta_{\pm}(\tau)=det[1-G_{\pm}]^{\infty}_{\tau}, \quad \tau=\pi t
\end{equation}

with the kernels $G_{\pm}$ defined by Eq. (\ref{eq101})

\begin{equation}\label{eq101}
G_{\pm}(x,y)=\delta(x-y)-\int^{\infty}_{0} \Psi_2(\mathbf{k,x}) \Psi_2(\mathbf{k,y}) dk
\end{equation}

Using the completeness relation,

\begin{equation}\label{eq102}
\delta(x-y)=\int^{\infty}_{0} j\left(k\left(x\pm \frac{1}{2}\right)\right)j\left(k\left(y\pm \frac{1}{2}\right)\right) dk
\end{equation}

then one can write Eq. (\ref{eq101}) to the form of Eq. (\ref{eq102})

\begin{align}\label{eq103}
&G_{\pm}(x,y)=\frac{1}{2}\int^{\infty}_{0} Re[(1-\exp(\mp i\theta(k)))h\left(k\left(x\pm \frac{1}{2}\right)\right) \nonumber \\
& \cdot h\left(k\left(y\pm \frac{1}{2}\right)\right)]dk  \nonumber \\
&=\frac{1}{4}\left[\int^{\infty}_{0}-\int^{0}_{-\infty}\right](1-\exp(\mp i\theta(k)))h\left(k\left(x\pm \frac{1}{2}\right)\right)\nonumber \\
& \cdot h\left(k\left(y\pm \frac{1}{2}\right)\right)dk
\end{align}

Here the function Eq. (\ref{eq104}) is analytic in the upper half-plane with a cut from ($+i$) to ($+i\infty$). 

\begin{equation}\label{eq104}
\exp(\mp i\theta(k))=\exp(\mp ik)(1\pm ik)(k^2+1)^{-1/2}
\end{equation}

\begin{Even Case}
	\emph{when $V_1=W_+$}
	
	In the even case, the exponential growth of Eq. (\ref{eq104}) is compensated by the exponential decrease of the Hankel functions in Eq. (\ref{eq103}) for all positive $x$ and $y$. 
\end{Even Case}

\begin{Odd Case}
	\emph{when $V_1=W_+$}
	
	In the odd case, Eq. (\ref{eq104}) decreases exponentially in the upper half-plane, but we must require $x,y>1/2$ so that the term in Eq. (\ref{eq103}) not involving Eq. (\ref{eq104}) has an exponential decrease. 
\end{Odd Case}

With this prescription, we may move the path of integration in both parts of Eq. (\ref{eq103}) to the positive imaginary axis by writing $k=iz$. The terms involving Eq. (\ref{eq104}) cancel along with the cut, and we are left with

\begin{align}\label{eq105}
&G_{\pm}(x,y)=\frac{2}{\pi^2}\left[\int^{\infty}_{0}dz-\int^{1}_{0}\exp(\pm \varphi(z))dz \right]\nonumber \\
& \cdot k\left(k\left(x\pm \frac{1}{2}\right)\right)h\left(k\left(y\pm \frac{1}{2}\right)\right)dk
\end{align}

\begin{equation}\label{eq106}
\varphi(z)=z-{\rm arctanh}(z)
\end{equation}

And the series expansion Eq. (\ref{eq107}) converges absolutely for all positive $\tau$ in the even case, and at least for Eq. (\ref{eq108}) in the odd case. 

\begin{equation}\label{eq107}
\log \Delta_{\pm}(\tau)=tr[\log (1-G_{\pm})]^{\infty}_{\tau}=-\sum^{\infty}_{1}n^{-1}tr[(G_{\pm})^n]^{\infty}_{\tau}
\end{equation}

\begin{equation}\label{eq108}
\tau > \frac{1}{2}+(4\pi)^{-1}
\end{equation}

The Eq. (\ref{eq99}) with Eq. (\ref{eq105}) and Eq. (\ref{eq107}) provides a practical method for computing the potentials $W_{\pm}(\tau)$, either by using the series expansion or by finding numerically the eigenvalues of the kernels $G_{\pm}$.

The relations Eq. (\ref{eq35}) and Eq. (\ref{eq99}), connecting $W_{\pm}(\tau)$ with $F_{\pm}(\tau)$ and $\Delta_{\pm}(\tau)$, have the consequence that the quantity

\begin{equation}\label{eq109}
\log F_{\pm}(t)+\frac{1}{4}\tau^2+\frac{1}{8}\log \left|\tau\pm\frac{1}{2}\right|-\log \Delta_{\pm}(\tau), \quad \tau=\pi t
\end{equation}

is a linear function of $t$. But we know that as $t\rightarrow\infty$ the behavior of log$F_{\pm}$(t) is governed by Eq. (\ref{eq88}), while $\log \Delta_{\pm}(\tau)$ tends to zero. The asymptotic Eq. (\ref{eq88}) can, therefore, be replaced by the identity

\begin{align}\label{eq110}
&\log F_{\pm}(t)=\nonumber \\
&-\frac{1}{4}\tau^2\mp\frac{1}{2}\tau-\frac{1}{8}\log \left|\tau\pm\frac{1}{2}\right|\pm\frac{1}{4}\log 2+\alpha+\log \Delta_{\pm}(\tau)
\end{align}

with

\begin{equation}\label{eq111}
\alpha=\frac{1}{24}\log 2+\frac{3}{2}\xi'(-1)
\end{equation}

\begin{equation}\label{eq112}
\log \Delta_{\pm}(\tau)=\sum(\pm1)^m a_m\left(\tau\pm\frac{1}{2}\right)^{-m}
\end{equation}

Then Eq. (\ref{eq35}) become Eq. (\ref{eq113}). Here, $t'=\pi t\pm\frac{1}{2}$

\begin{align}\label{eq113}
&V(t')=\left(\frac{2m}{\hbar^2}\right)W_{\pm}(\tau)\nonumber \\
&=\left(\frac{2m}{\hbar^2}\right)\left[-\frac{1}{4}(t')^{-2}-2\sum m(m+1)(\pm1)^m a_m (t')^{-m-2}\right]
\end{align}

So this shows we can uniquely construct the potential $V$ from the wave function.
Therefore we can prove the one-to-one relation between $V(x)$ and $\Psi(\mathbf{k,x})$. \qed



\end{document}